# Scalable Nernst thermoelectric power using a coiled Galfenol wire

Zihao Yang[1], Emilio A. Codecido[2], Jason Marquez[3], Yuanhua Zheng[4], Joseph P. Heremans[2,4] and Roberto C. Myers[1,3,4a)]

[1] Electrical and Computer Engineering, The Ohio State University, Columbus, Ohio

[2] Physics, The Ohio State University, Columbus, Ohio

[3] Materials Science and Engineering, The Ohio State University, Columbus, Ohio

[4] Mechanical and Aerospace Engineering, The Ohio State University, Columbus, Ohio

Email: myers.1079@osu.edu , Web site: http://myersgroup.engineering.osu.edu

**The Nernst thermopower usually is considered far too weak in most metals for waste heat recovery. However, its transverse orientation gives it an advantage over the Seebeck effect on non-flat surfaces. Here, we experimentally demonstrate the scalable generation of a Nernst voltage in an air-cooled metal wire coiled around a hot cylinder. In this geometry, a radial temperature gradient generates an azimuthal electric field in the coil. A Galfenol ($Fe_{0.85}Ga_{0.15}$) wire is wrapped around a cartridge heater, and the voltage drop across the wire is measured as a function of axial magnetic field. As expected, the Nernst voltage scales linearly with the length of the wire. Based on heat conduction and fluid dynamic equations, finite-element method is used to calculate the temperature gradient across the Galfenol wire and determine the Nernst coefficient. A giant Nernst coefficient of -2.6 µV/KT at room temperature is estimated, in agreement with measurements on bulk Galfenol. We expect that the giant Nernst effect in Galfenol arises from its magnetostriction, presumably through enhanced magnon-phonon coupling. Our results demonstrate the feasibility of a transverse thermoelectric generator capable of scalable output power from non-flat heat sources.**

Thermoelectric power generation is considered an environmentally friendly approach to convert waste heat into electrical energy. The conventional way is to take advantage of the Seebeck effect, where an electric field is generated longitudinally along a thermal gradient. On the other hand, a transverse thermoelectric effect, called the Nernst effect, generates an electric field perpendicular to the plane between an applied temperature gradient and magnetic field. Despite the weaker nature of the Nernst effect compared to the Seebeck effect, the Nernst effect offers several advantages and flexibilities for thermopower generation [1]. More specifically, due to the orthogonality of the Nernst electric field to the thermal gradient, the Nernst thermopower can be increased linearly with the thermoelectric generator length without introducing the conventional thermopile structure (n- and p-type semiconductor structures connected in series). This is specifically useful for cylindrical heating geometries, where a device originally proposed by Norwood [2] can be implemented. This device consists of a wire wrapped around a hot cylinder to produce a Nernst voltage that scales linearly with the wire length, i.e. the number of turns. As pointed out by Norwood [2], "the value of a device of this type is that the heat pumping material has no metal-semiconductor junctions and there are no heat leakage paths (neglecting end effects) except through the material." The scalable Nernst thermopower generation possible in the coiled wire geometry also provides a more easily manufactured and economical approach than the conventional Seebeck thermoelectric generator. Thus, the Nernst effect offers an attractive alternative for thermoelectric waste heat recovery on non-flat surfaces, such as hot exhaust pipes.

In magnetic materials, the total Nernst effect consists of ordinary ($\vec{E}_{ONE}$) and anomalous ($\vec{E}_{ANE}$) components:

$$\vec{E}_{ONE} = N_{ONE}\mu_0(\vec{H} \times \vec{\nabla}T), \quad [1]$$

$$\vec{E}_{ANE} = N_{ANE}\mu_0(\vec{M} \times \vec{\nabla}T), \quad [2]$$

where $N_{ONE}$ and $N_{ANE}$ are the ordinary and anomalous Nernst coefficients respectively, $\mu_0$ is the vacuum permeability, $\vec{H}$ is the applied field, and $\vec{M}$ is the magnetization. Sakuraba recently evaluated the potential performance of a thermoelectric generator in a coil geometry and showed that in order to reach practical power densities useful for waste heat recovery, larger Nernst coefficients are necessary compared to those typically found in metals [1]. To that end, phonon drag had been shown to enhance the Nernst coefficient at low temperatures in certain material systems [3-4]. More recently, Watzman *et al.* reported that the anomalous Nernst coefficient also benefits from magnon drag as shown in single crystal Fe where $N_{ANE}$ is dominated by magnon drag up to 200 K [5]. In this study, we propose using a magnetostrictive metallic metal, Galfenol ($Fe_{100-x}Ga_x$), as a pathway to increase the power factor due to its enhanced Nernst coefficient from phonon and magnon coupling, as well as its large intrinsic electrical conductivity. We demonstrate Nernst voltage generation in a coiled Galfenol wire, characterize its Nernst coefficient, and offer an experimental proof-of-concept for a novel and scalable thermoelectric generator ideal for cylindrical heaters.

Galfenol (FeGa) represents a category of mechanically robust, high strength, and low cost magnetostrictive alloys. The magnetostriction is shown to be as high as ~ 886 ppm for $Fe_{83}Ga_{17}$ with Tb doping [6]. It was shown that Galfenol exhibits higher magnetostriction with increasing Ga content up to ~ 20 at% Ga; however, Galfenol also exhibits the ductile-to-brittle transition at 15 at% Ga [7-9]. Therefore, ductile Galfenol wires with composition of 85 at% Fe and 15 at% Ga obtained commercially from Etrema Products INC are used in this experiment. The wire diameter is ~ 508 μm and the fabrication process is similar to Ref. [9].

The Galfenol wires are varnished with Glyptal 1201 enamel or inserted into a polytetrafluoroethylene (PTFE) shrink tube, resulting in a ∼ 70 or 250 μm thick layer for electrical insulation, then wrapped around a cylindrical cartridge heater as shown in . 1(a). In cylindrical coordinates, the heater generates a radial temperature gradient, $\vec{\nabla}T = \nabla T_r \hat{r}$, throughout the Galfenol wire and a magnetic field in the axial direction, $\vec{H} = H_z \hat{z}$, is applied. Following Eq. 1 and 2, an azimuthal electric field is produced at every point along the coil,

$$\vec{E}_{NE} = N_{ONE}\mu_0 H_z \nabla T_r \hat{\varphi} + N_{ANE}\mu_0 M_z \nabla T_r \hat{\varphi},$$

therefore a voltage drop, $V$, across the wire is produced that scales with the wire length. Glyptal enamel is used for the demonstration of large voltage generation due to its higher working temperature and thinner insulation, while PTFE is used for the estimation of the Nernst coefficient from simulations due to its better known physical parameters. The Nernst voltage is measured using a custom-made system including an electromagnet from SES Instruments and Keithley 2700.

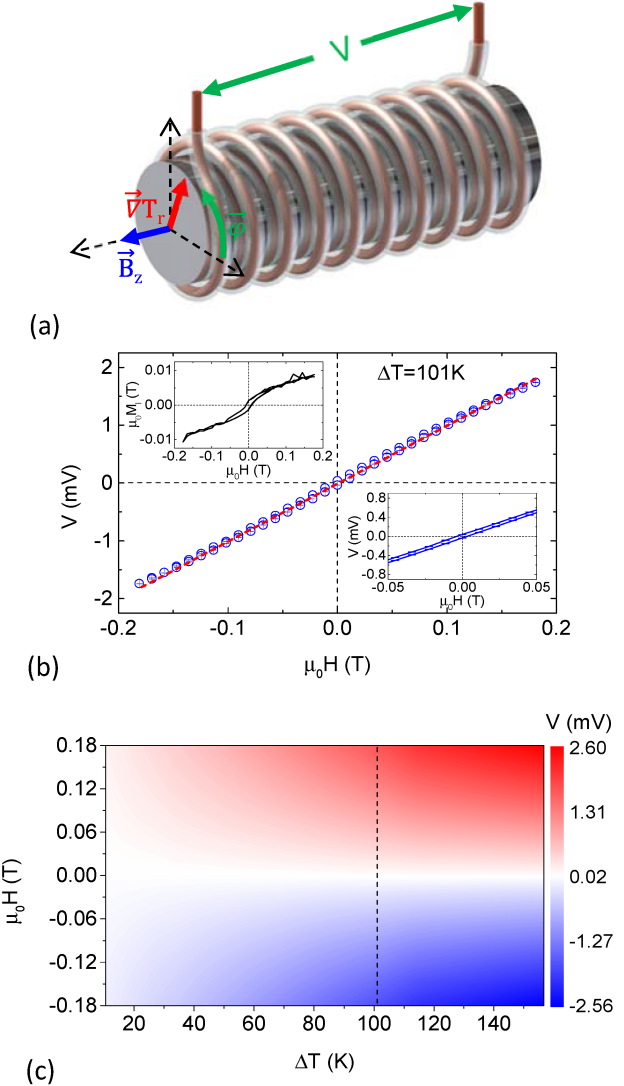

Fig. 1(a) Schematic diagram of the device setup. The Galfenol wire, cartridge heater, and electrical insulation material are shown as brown solid coil, gray solid cylinder, and semitransparent tube respectively. (b) Magnetic field dependent Nernst voltage at a radial temperature difference of ΔT = 101 K. Top left inset: local magnetization (M$_l$) of the Galfenol wire at room temperature. Bottom right inset: Nernst voltage in low field region. (c) 2D contour map of Nernst voltage as a function of the magnetic field and radial temperature difference.

Figure 1(b) plots the raw magnetic field dependent Nernst voltage, *V*, (blue open circle) generated in a coiled Galfenol wire (~ 75 cm) coated with enamel, under radial temperature drop of ~ 101 K with the magnetic field swept in a hysteretic manner. The Nernst voltage is almost entirely linear with magnetic field demonstrating a slope of ~ 10 mV/T and a maximum voltage of ~ 1.74 mV at a magnetic field of 0.18 T and an input heating power density of 5.2 W/cm$^2$. The Nernst voltage at low magnetic field ($\pm$ 0.05 T) is shown in Fig. 1(b), bottom right inset, after subtracting the background voltage at zero field (~ 30 µV). This small background voltage could result from the Seebeck effect due to a small temperature difference between slightly misaligned electrical leads. An open hysteresis loop is observed in the Nernst voltage, as shown in Fig. 1(b), bottom right inset, indicative of the anomalous contribution (M dependent) to the Nernst effect. Qualitatively identical behavior is observed in samples with different lengths.

Accurate measurement of the magnetization of the coil device is unfeasible using standard magnetometry methods due to the large sample size and coil geometry. Instead, we measure the local total field ($B_l$) for qualitative magnetization behavior by placing a digital teslameter in the $r - \varphi$ plane (perpendicular to $\vec{B}_z$) in close proximity to the coil end. The local applied field ($H_l$) is obtained with the coil removed. The local magnetization ($M_l$) is therefore determined by subtracting the local applied field from the local total field, $\mu_0 M_l = (B_l - \mu_0 H_l)$, plotted in the top left inset of Fig. 1(b). The $M_l$-$H_l$ behavior reveals a non-zero remanence and $M_l$ does not saturate under the maximum applied field of 0.18 T. We ascribe this hard axis behavior to the shape anisotropy due to the coiled wire geometry. Although it is expected that the anomalous Nernst voltage dominates in ferromagnets before reaching saturation magnetization, the difference in linearity between the Nernst and magnetization data suggest contributions from both ordinary and anomalous Nernst effects. The zero-field corrected Nernst voltage, as a function of magnetic field and radial temperature gradient, is shown in Fig. 1(c). Line cuts along the y-axis at a fixed radial temperature gradient show a linear relationship between the Nernst voltage and magnetic field (blacked dashed line represents the data shown in Fig. 1(b)), where the slope increases from 84.3 µV/T to 15 mV/T with an increasing temperature drop from ~ 10.6 to ~ 154.1 K. The Nernst thermopower (NB) is defined as $\alpha_{r\varphi z} \equiv \frac{E_\varphi}{\nabla T_r}|_{\mu_0 H_z}$ (similar as in Ref. [5]) and can be estimated using the slope of the line cuts along the x-axis and thermal conductivities of the enamel [10] and Galfenol [see Tab. S1], where a magnitude of ~ 0.33 µV/K is found at a magnetic field of 0.18 T near room temperature. The Galfenol coil with PTFE insulation yields a similar value of ~ 0.4 µV/K based on the temperature simulation as discussed below.

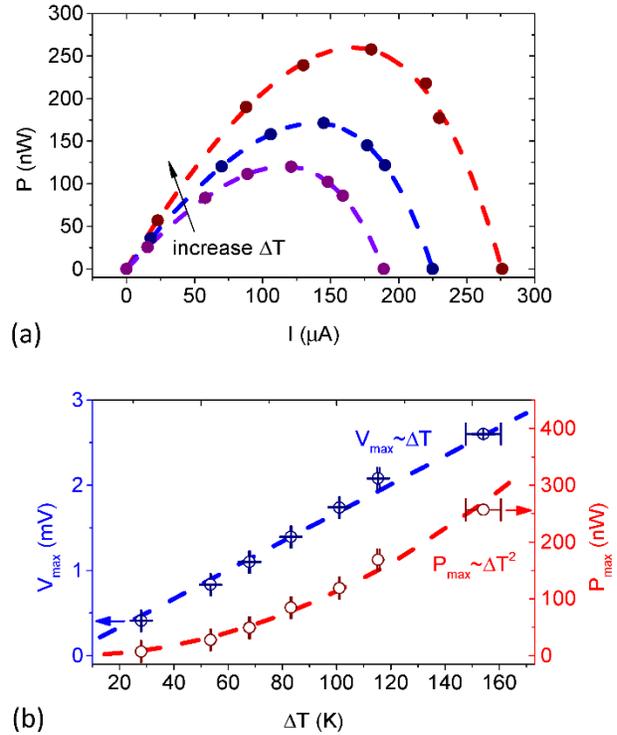

The output power from the enamel coated Galfenol coil is measured by connecting the Nernst coil to external load resistors and measuring the current and voltage across the load resistors. An example of the measured output power as a function of the current flowing through the external resistor at different heater power is shown in Fig. 2a. The shape of the power curves measured at a ΔT of 101 K, 115.3 K and 154.1 K closely resembles each other. The maximum output power is determined by interpolating between the measured values where a maximum of ~ 260 nW can be found at a maximum temperature drop of ~ 154.1 K across the wire. The maximum Nernst voltage and output power as a function of measured temperature drop are plotted as open circles in Fig. 2b. The maximum Nernst voltage, taken from the line cut along the x-axis at a magnetic field of 0.18 T in Fig. 1c., increases proportional to the temperature difference as shown by the linear fitting. The output power calculated based on the thermal circuit of the device gives $P_{out} = \frac{\pi N^2 B^2 L_{GaFe}}{16 \rho_{GaFe}} \left[ \frac{1}{1 + 2 \frac{d_{Glyptal} \kappa_{Glyptal}}{d_{GaFe} \kappa_{GaFe}}} \right]^2 \Delta T^2$. Here, $N$ and $B$ are the Nernst coefficient and magnetic field, $\rho$ and $\kappa$ are the electrical resistivity and thermal conductivity, $d$ and $L$ are the diameter and length and $\Delta T$ is the temperature difference across the enamel coated Galfenol. Therefore the maximum output power is fitted using $P_{max} \sim A \cdot \Delta T^2$ and a general good agreement with the measured values is achieved. This also implies that the prefactor $A$ has a rather small dependence on the temperature. At the maximum temperature, the much larger error bar in the temperature measurement is due to the near-breakdown of the enamel insulation.

Fig. 2(a) The output power from the Galfenol coil with Glyptal as a function of the current flow through the load resistor at different temperatures drops across the wire. (b) The maximum voltage and output power with linear (~ ΔT) and parabolic fit (~ ΔT²) as a function of temperature drop across the wire.

A chief advantage of Nernst power generation in the coil geometry is its scalability with wire length ($L$). To demonstrate this idea, the background corrected Nernst voltage at a fixed temperature gradient is measured with $L$ = 12, 35, and 59 cm on three Galfenol coil devices with PTFE insulation, as shown in Fig. 3(a). An increase in Nernst voltage at any given magnetic field is observed by increasing the wire length. This is more clearly shown in Fig. 3(b), where the length-dependent Nernst voltage is plotted directly. The Nernst voltage is linear with $L$ at different thermal gradients, demonstrating voltage scalability in the Nernst coil.

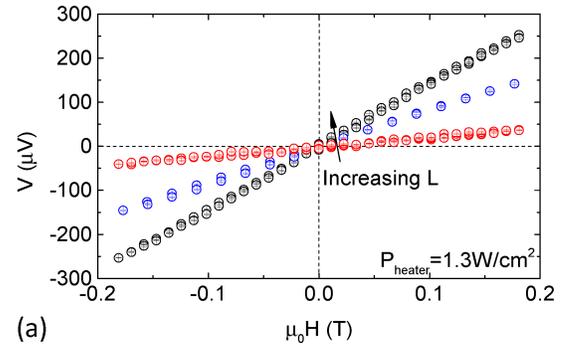

(a)

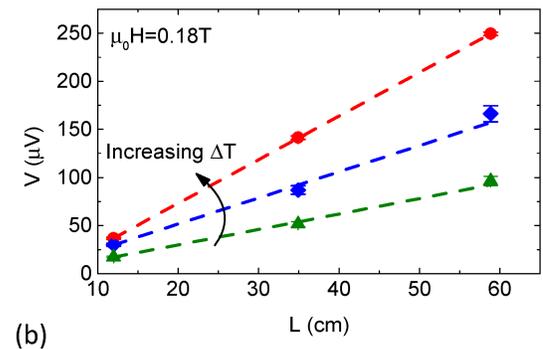

(b)

Fig. 3(a) Nernst voltage as a function of magnetic field at a heater power density of 1.3 W/cm$^2$ with different Galfenol wire lengths. (b) Nernst voltage as a function of wire length with different temperature differences.

This wire geometry also serves as an alternative way to measure the Nernst coefficient for the materials that are already in the wire form. However, since polymer based materials are used to isolate the Galfenol wire electrically from the cartridge heater, most of the temperature drop is across the Glyptal or PTFE due to their low thermal conductivities. This makes a direct temperature measurement across the Galfenol wire both impractical, due to its geometry, and unreliable, due to the fraction of a degree temperature drop expected across it. Additionally, an external air flow directed towards our setup was introduced to serve as active cooling and aid temperature stabilization, making the heat flow azimuthally dependent and difficult to estimate. Therefore, finite-element method software COMSOL [11] is used to simulate the temperature profile in the Galfenol devices with PTFE insulation using the parameters from [12-18]. The details of the thermal modeling and input parameters are discussed in the supplementary materials.

The Nernst coefficient defined as $N \equiv \frac{\partial \alpha_{r\varphi z}(\mu_0 H_Z)}{\partial \mu_0 H_z}$ (similar as in Ref. [5]) is computed using the thermal modeling results from the supplementary materials and is shown as a function of temperature in Fig. 4 (black triangles). The Nernst coefficient of the Galfenol wire is found to be -2.6 µV/KT near room temperature and relatively constantly up to 315 K. This is compared to the results (red stars) from another independent measurement on a 6 × 2 × 1 mm bulk polycrystalline Galfenol with a Ga content of 18.4% obtained from the same vendor measured between ± 1.4 T in a custom-made cryostat. The measurement setup is similar to the one used in Ref. [19] and is also used for Galfenol thermal conductivity and specific heat measurements as discussed in the supplementary material. The Nernst thermopower of the bulk sample, $\alpha_{xyz}$, at 82 K and 300 K shown in Fig. 4 inset does not fully saturate up to 1.4 T and is similar to the coiled wire measurement. The Nernst coefficient in the Galfenol wire ($Ga_{15}Fe_{85}$) is found to be slightly lower (by ~ 20%) than the one measured in the bulk sample ($Ga_{18.4}Fe_{81.6}$) at around room temperature possibly due to the difference in their magnetostriction. To the best of our knowledge, this is the first measurement on the Nernst coefficient of Galfenol. This value is higher than the reported $N$ of 0.189 μV/KT in Ref. [20], 0.4 μV/KT in Ref. [5], and 0.56 μV/KT in Ref. [21] for single crystal $Fe_3O_4$ bulk, single crystal Fe bulk, and epitaxial FePt thin film, respectively, near room temperature. The Nernst coefficients in Ref. [20] and [21] are calculated based on the saturation magnetization ($\mu_0 M_s$) of the sample and are converted to saturation magnetic field ($\mu_0 H_s$) to compare to our result. The Nernst coefficient in the bulk Galfenol sample continuously increases in magnitude from 80 to 300 K. The physical origin of the temperature dependence requires further investigation.

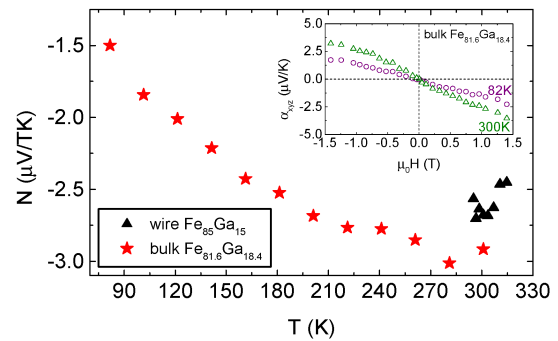

Fig. 4 Comparison of the Nernst coefficient between bulk ($Fe_{81.6}Ga_{18.4}$) and wire ($Fe_{85}Ga_{15}$) samples. Inset: Nernst thermal power as a function of magnetic field measured in bulk sample at 82 K and 300 K.

In conclusion, a coiled wire wrapped around a cylindrical heater is experimentally demonstrated for the first time as a proof of concept for an easily scalable thermoelectric generator utilizing the Nernst effect. The thermovoltage scales linearly with coiled wire length, providing a linearly scalable output power for proper load impedances. In addition, a giant Nernst coefficient of ~ -2.6 to -3 μV/KT at room temperature is measured in Galfenol (FeGa alloy). Other magnetostrictive materials, like Galfenol, are expected to exhibit stronger magnon-phonon coupling, and therefore through magnon-electron drag, lead to an increased Nernst effect. This offers one possible explanation for the observed giant Nernst effect in Galfenol compared with pure Fe. Upon discovering new magnetostrictive materials that possess both higher Nernst coefficient and remnant magnetization as well as employing higher thermal conductivity insulation, the Nernst coil generator may become a feasible device for waste heat recovery. A new model for the Nernst effect including strongly hybridized magnon-phonon excitations (or magnon polarons) is needed to reveal the physical origin of the high Nernst coefficient and the temperature dependence of magnetostrictive materials.

This work was supported by the Army Research Office MURI W911NF-14-1-0016 and by the NSF-REU program of the Center for Emergent Materials: an NSF MRSEC (DMR-1420451). The authors thank Prof. Marcelo Dapino for providing the bulk Galfenol samples. The authors thank John Jamison for experimental support and valuable discussions.

# Finite-element method modeling of the temperature profile in Galfenol with PTFE shrink tube

Finite-element method software COMSOL [10] is used to simulate the temperature profile in the Galfenol wire with PTFE shrink tube. Fig. S1(a) shows the simulated surface temperature profile in our setup with a wire of $L$ = 12 cm, including the wind speed along the r-z plane ($\varphi$ = 0). The simulation consisted of an inlet air flow, which was measured to be ~ 7 m/s. In this geometry, 7 m/s corresponds to a Reynolds number in the turbulent regime; therefore, we used a low Reynolds k-ε turbulent model to simulate the air flow profile. This model shares the advantages of the k-ε model, but does not include wall functions, thereby solving for the air flow everywhere. This is necessary for a proper heat transfer and convective cooling coupling near the boundaries. The temperature profile was simulated by combining the convective cooling due to the resulting air flow profile and inputting the cartridge heating power calculated using Joule's law. A thermocouple was attached to the surface of the coil, serving as a temperature point to cross-check the simulations. The thermal conductivity ($\kappa$), specific heat ($C_p$), density ($\rho$), ratio of specific heat ($\gamma$), and dynamic viscosity ($\mu$) used for these simulations are shown in Table S1. These parameters are assumed to be temperature independent due to the relative small absolute temperature change in the measurement. The thermal properties of Galfenol are measured from a separate custom-made cryostat setup, as in Ref. [1], on a bulk sample from the same vendor with a very slight difference in composition. The density of Galfenol is estimated based on the atomic percent of Fe (85%) and Ga (15%). Fig. S1(b) shows a good agreement between measured and simulated temperature at the same point (gray pentagon in Fig. S1(a)) on the coil at different cartridge heating powers. Simulations in Fig. S1(c) show the average temperature (T$_{average}$) and

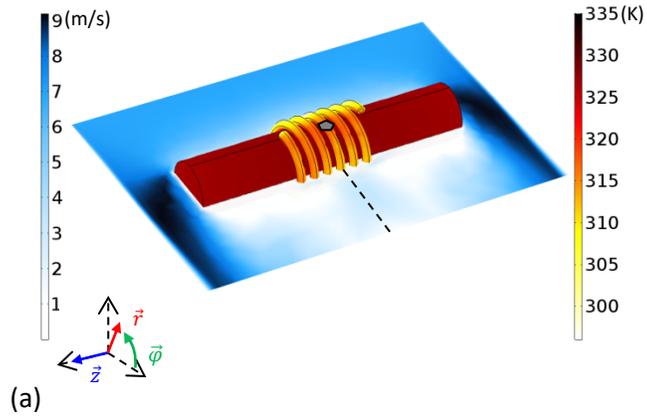

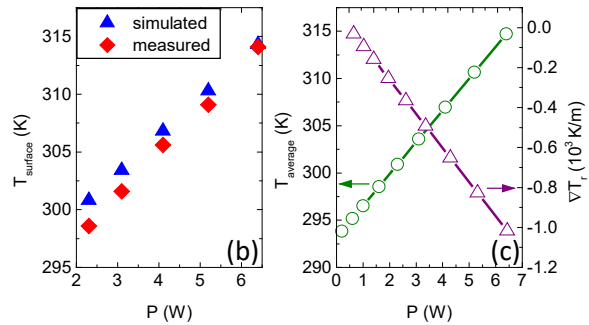

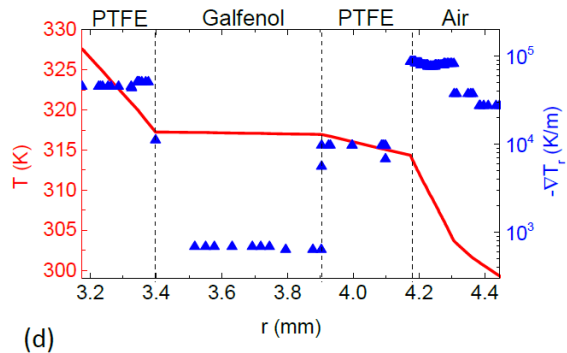

Fig. S1 (a) 3D plot of surface temperature and wind speed simulated by COMSOL. (b) Comparison between measured and simulated surface temperature on gray spot in Fig. S1a. (c) Average temperature and temperature gradient for the Galfenol wire as a function of cartridge heating power. (d) Temperature and temperature gradient profile along the black dotted line in Fig. S1a at heater power of 6.4 W.

average radial temperature gradient ($\nabla T_r$) in the Galfenol wire as a function of input heating power.

These values are later used to calculate the Nernst coefficient of the Galfenol wire. Fig. S1(d) plots the simulated radial temperature profile and temperature gradient across the coil device along the black dotted line in Fig. S1(a). At an input power of 6.4 W, a radial temperature drop of ~ 0.3 K is produced across the Galfenol while, as expected, a more significant temperature drop is found in PTFE and adjacent air layers.

Table S1. Material parameters used in simulation

|  | $\kappa$ (W/(m·K)) | $C_p$ (J/(kg·K)) | $\rho$ (kg/m$^3$) | $\gamma$ | $\mu$ (Pa·s) |
|---|---|---|---|---|---|
| Air | 0.024[2] | 1005[2] | 1.2[3] | 1.4[3] | $2 \times 10^{-5}$[4] |
| Heater (copper) | 387[5] | 385[5] | 8930[5] | - | - |
| Shrink tube (PTFE) | 0.25[6] | 1300[7] | 2200[8] | - | - |
| Galfenol | 17 | 437 | 7329 | - | - |